# Design & Deploy Web 2.0 enable services over Next Generation Network Platform

Dr. Kamaljit I. Lakhtaria[*], Dhinaharan Nagamalai[#]
[*]Atmiya Institute of Technology & Science, Rajkot, Gujarat, INDIA
Email: kamaljit.ilakhtaria@gmail.com
[#]Wireilla Net Solutions PTY LTD, Australia

**Abstract.** The Next Generation Networks (NGN) aims to integrate for IP-based telecom infrastructures and provide most advance & high speed emerging value added services. NGN capable to provide higher innovative services, these services will able to integrate communication and Web service into a single platform. IP Multimedia Subsystem, a NGN leading technology, enables a variety of NGN-compliant communications services to interoperate while being accessed through different kinds of access networks, preferably broadband. IMS–NGN services essential by both consumer and corporate users are by now used to access services, even communications services through the web and web-based communities and social networks, It is key for success of IMS-based services to be provided with efficient web access, so users can benefit from those new services by using web-based applications and user interfaces, not only NGN-IMS User Equipments and SIP protocol. Many Service are under planning which provided only under convergence of IMS & Web 2.0. Convergence between Web 2.0 and NGN-IMS creates and serves new invented innovative, entertainment and information appealing as well as user centric services and applications. These services merge features from WWW and Communication worlds. On the one hand, interactivity, ubiquity, social orientation, user participation and content generation, etc. are relevant characteristics coming from Web 2.0 services. Parallel IMS enables services including multimedia telephony, media sharing (video-audio), instant messaging with presence and context, online directory, etc. all of them applicable to mobile, fixed or convergent telecom networks. With this paper, this paper brings out the benefits of adopting web 2.0 technologies for telecom services. As the services are today mainly driven by the user's needs, and proposed the concept of unique customizable service interface.

## 1. Introduction

The essential inspiration of Next Generation Networks (NGNs) [1] is to carry all types of service on a single packet-based network. This 'network convergence' allow operators to save money by having to maintain only one network platform, and to provide new services that combine different types of data. NGNs are more versatile than traditional networks because they do not have to be physically upgraded to support new types of service. The network simply transports data, while services are controlled by software on computers that can be located anywhere. This means that third parties can easily launch new services, not just the network operators themselves.

NGNs are based on IP, like the Internet, but they build in features that the Internet does not have, such as the ability to guarantee a certain quality of service and level of security. For example nowadays communications target to transmit a variety of of services. Those are classical telephony also the Internet traffic, data transmission, radio and television broadcasting etc. Consequently, various transmission media are used as metal and fiber cables, and microwave, millimeter wave, and optical free space communication links. Most definitions of NGN also include the principle of 'nomadicity'. This means that a user can access personal network services from different locations using a range of devices such as fixed-line phones, mobile phones and computers.

### 1.1 NGN-IMS Architecture

The idea of Next Generation Network (NGN) [2] is services such as voice, data and all sorts of multimedia or communication services to be transported into single infrastructure, which will be Internet Protocol (IP) based. IP Multimedia Subsystem (IMS) is one of the underlying technology components of Next Generation Network which services should be independent to any fixed or mobile networks. The IMS is introduced in order to provide converged services in IP-based network. The advent of IP Multimedia Subsystem (IMS) leads fixed and mobile communication services to convergent evolution and improves the user experience. The architecture of IMS is designed for fast deployment of services, assuring end-to-end Quality of Service and integration of different services.

Nowadays there are many services deployed over IP-based network such as instant messaging, VOIP, presence and video sharing. Service providers deploy those services in uncoordinated way. A manageable framework like IMS is required to manage those services and provide flexible integration and deployment for service providers.

With the standardization of the IMS architecture [1][2][3], the service development methods of telecom operators become more and more similar to the web development methods. Indeed, service development environment of telecom operators is based on reusability of the basic enablers (e.g. presence, messaging) which is very similar to a service oriented architecture (SOA) approach [4] that has proved its usefulness in the decade. Moreover, several operators have even opened their network through OSA/ParlayX [5] web services to facilitate the development of new telecom services. However, the promised innovative services take a long time to appear because of difficulties to manage the real time applications in the web environment. The web community has otherwise gained in experience of real time applications (e.g. googleTalk, and webMessenger) that uses web 2.0 [6] technologies such as AJAX [7][8]. Moreover, innovative applications have appeared on the public web, such a web aggregators, mashups, and social networking. These applications are characterized by the aggregation of services, sharing, participation and personalization.

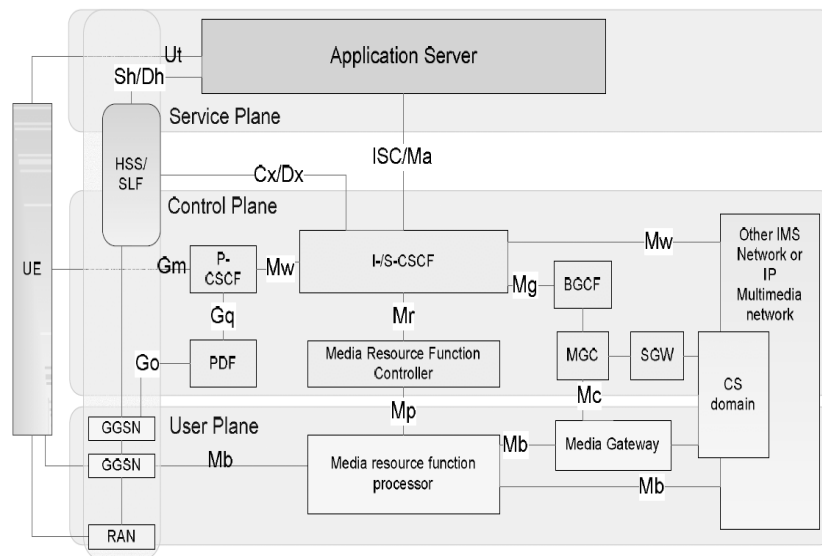

**Figure 1.** IMS Layered Architecture [3]

Web aggregators (like Netvibes [9] and iGoogle [10]) are applications that give access to many web feeds (data format used to provide users with dynamic content such as news and weather) of many providers from a single web page; these feeds are displayed as independent blocs in the page (e.g. weather feeds, sport feeds, political news feeds). The goal of this paper is to show how telecom operators can benefit from web technologies to provide a unified point of access to user's services (web information and telecom services). We consider the unified interface concept as a first step toward those innovative services. Therefore, for this analysis work, research have implemented dashboard that blends both telecom real-time services and web information services on a single web page. This enables the user to access, to monitor, and to use all its preferred services simultaneously. Presented solution is implemented at the presentation layer. This is a definitely a new approach for the convergence of the telecom domain and the web domain. The real-time issues are handled using web 2.0 [6] technologies.

The rest of this paper is organized as, Section 2 Demonstrate through examples the added value of services aggregators for both the operator and the end user. While Section 3 Befits with Value Added Services with Web Aggregator implementation work discussion continues to Section 4 & 5 briefs Real-time Web aggregator implementation demonstration with their Function & Architectural descriptions. Section 6 discusses Several Implementation Issues like Security, trade-off notion of this Convergence model with respective future work path in Conclusion.

### 1.2 Web 2.0 Environments for NGN

Web 2.0, the next generation networking is all about making the web as platform, harnessing collective intelligence, lightweight and meshup programming models, providing rich user experiences, and approaching application development differently. With all these characteristics, web 2.0 is impacting various applications that have been available in traditional web such as personal websites, sharing and accessing data, connecting with other people over the web, etc. It is also having impact on the capabilities of different platforms/devices such as personal computers, mobile devices and other handheld devices that access such web applications.

Specifically, the present mobile devices with their limited infrastructure restrict the users' [20] interaction and limit their ability to interact with the web applications in a full-fledged manner.

The improvement and enrichment to mobile communication technology and its underlying structure to access the advanced feature of the web is being referred to as mobile web 2.0 or simply mobile 2.0. Mobile web 2.0, with myriad of possibilities to offer, is changing the way people interact with the web, pushing their limits up to levels that they could only imagine in the past using mobile devices. It promises to transcend beyond the limitations of the previous generation of mobile devices, enhancing and enriching the overall experience of using the web through these tiny devices.

Mobile web 2.0 is about leveraging the power of the web, integrating the web services and their features on the mobile platform, and providing the users the rich experience they enjoy on desktop applications. Mobile web 2.0 focuses on harnessing the strength and capabilities of the applications supported by web 2.0 and extend them to the mobile platform, making it more powerful and usable. The concept of mobile 2.0 does not limit itself to only a traditional handheld device. In the future, mobile 2.0 devices would be so tiny that they could be implanted in our clothing, physical environment, accessories, and body parts, enabling us to instantaneously connect to other people laden with such devices, interact with the web subconsciously and perform other functions now only possible through desktops. This coincides with the W3C vision for the web—"Universal Web Access: The Web Anywhere, for everyone, at anytime, on everything" [21].

## 2. Related work

Web Aggregator word simply means "collector of multiple entitites", Customizable Web feed aggregators such as Netvibes [9] and iGoogle [10] provides the user with (1) the ability to access many feeds from a single web page, (2) the ability to integrate third party feeds, and (3) personalization capabilities.

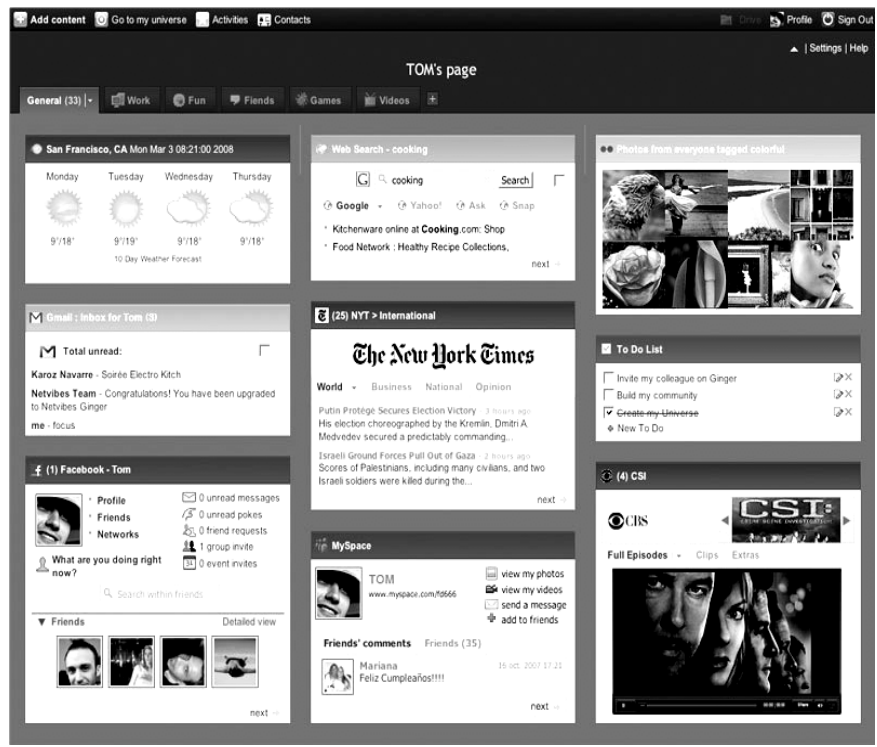

**Figure 2**: Netvibes web screen shot [9] Source: www.netvibes.com

### 2.1   To retrieve multiple fees at one Platform

Web aggregator simply called Module or widget sometime also called portal [11]. A Web aggregator mentioned with URL must concerned and serve with a presentation layer. While aggregator sends requests through its URLs of each module/widget including given user and user's request information as well. The widgets perform as independent performance of user's expected service out of all Web Pages. Each interaction regarding Aggregators' service deployment passes through business logic with AJAX. AJAX provides web framework to update as per business logic & followed by hosted on the server, but this all process not consume any need of page refreshing. Through AJAX this architecture keep modules independent each from others.

### 2.2 Easy integration with third party feeds & User Personalization

AJAX technologies enable aggregators to integrate modules given by different providers. Service providers as well as independent developers easily develop their own widgets. There are many ways to develop a widget. Actually, this depends on the target aggregator (whether this widget will be incorporated in iGoogle, Netvibes, or another aggregator). However, the universal widget API UWA [12] has gained a large community (Netvibes, iGoogle, Mac, Windows vista, Yahoo! Widgets, & Opera).

Moreover, W3C has initiated a standardization effort of widgets development API [13]. User able to retrieve single service from multiple time widgets there is no conflict between any Aggregator. Same way if any Aggregator service provider provides more than one fees, all fees can be presented in single web page. Web Aggregators able provide personalization; each user can personalize its aggregator by managing own modules into their Web pages. Above Figure 2 is a Netvibes [9] screenshot presented characteristics of feed aggregators.

### 2.3 IMS Architecture Support for Web 2.0 Integration

While mashing up an application the controller of the client tier coordinates several services ranging from Web Services to SIP based IMS services that implements diverse capabilities in the network. A service, which wants to be a part of the coordinated service, contacts the SC and asks it to create a coordination context. The SC will create a new instance of the coordination context and hand the context over to the requesting participant service. A coordination context is actually a data-structure that contains an activity identifier. The SC also passes the coordination context to the client tier for further correspondence with the mashup application. The requesting participant service will then ask the registration service of the SC to register its role in the coordinated activity. The registration service will register the role of the participating service in the activity. The participating service then might propagate the coordination context instance to other services that it would like to take part in the same activity.

## 3. Value Added Services with Web Aggregators

Web Aggregating services convergence into one web page able to provide higher benefits to user as well as service provides.

### 3.1 End users benefits

**Increasing interaction between independent services:** End user access there instantly. User able to retrieve more than one service. No need to navigate between services all service easily available at a single platform. User able to combine two services like select one entity from one feeds and uses that one another one as well. For example, while user uses soft phone through web feeds, any contact will retrieved as well as add to contact directory. This is main benefit that all service either provided by same provider or by different provider but once retrieved by user, then follow user's direction only. Same way using Google Map user directly fetch address and able to see location into single web platform.

**Easy access to third party service providers:** The user also able to retrieve services from third-party providers. Specially News Fees, Entertainment, Travel Information all managed by third party service providers only. But user easily access this all. Many developer / freelancer also provide customized Aggregators to particular services. These all services easily integrate web page.

**User authentication & Personalization:** User always worry for security with web Aggregators users access to their services from a single web page, the operator can perform a single authentication for all provided services. With this feature users does not requires to re-authenticate for each service. E.g. if the same operator, the user then, provides the phone service and the directory service performs a single authentication for both services. Users can personalize theirs with organizing their services into groups through tabs. They can also add, move, and delete services at the run time. All configurations (tabs configuration and modules configurations) are saved and retrieved at each disconnection and connection.

### 3.2 Operator advantages

Operators able to know its users, Communication service provider as well as Entertainment and other service provides able to know gadgets used chosen by their user. With this service provide enhance and upgrade their services. Making a widget as simple as possible is the slogan of the widgets developers; indeed a widget is supposed to perform a basic function, analogous to IMS enablers [16]. Telecom operators also develop Aggregators that will enhance basic communication related services like presence and IM. Single accessing interface enables the operator to manage a single authentication for all their services. This facilitates telecom operators in the management of theirs services.

## 4. Contributions and functional description

This observation made through implementation of unified dashboard using Internet over telecom services. Presented Unified dashboard belongs to a web page that will aggregate web services as well as telecom services. Web services consist for example in web search, map, and RSS feeds. And telecom services include for instance telephony, messaging services, and videoconferencing. Users can access to their services through the dashboard, by using an Internet connection and a web browser whatever the used device like mobile phone, PDA, laptop, and desktop PC. This converged dashboard able to manage user for derive multiple services from different providers.

As per shown in Figure 3, our implemented desktop able to serve Multiple Convergence service at once. Service with short description as following:

1. Dashboard displays User's Profile, while user log-in. User able to manage his profile as well as own setting to manage Dashboard.
2. Implemented Dashboard in short performs Speed dial, where user directly select their buddy list from dashboard and make call to them, in cell phone maximum up to 9 speed dial number assigned, but with Dashboard – Buddy listing user able to add more contacts to speed dial.
3. Using API, User able to retrieve any Gadget from Web Feed Provider, mostly able to retrieve from iGoogle, Netvibes etc., in this Dashboard News Feed Tag prepared to retrieve from any News using API.
4. More Communication Service like Voice Call, Voice message also presented with this Dashboard, While User perform operation with third party Gadget/API using Feeds at that time, User Identity must provided separately to that service (e.g. in Fig. 3 Call Wave Feeds).
5. Same like Web, User able to add more Feeds for Information, Entertainment, Fun also. In this Dashboard, Picture tag is fetched using Feeds over Photo Shot Gadgets, further more user also able to connect through Video Sharing dashboard.
6. Dashboard easily integrates many different applications to a single user interface platform. Once user authentication process completes, user selected gadgets works to retrieve data into own page. For this concept the service provided with XHTML / AJAX technology.
7. Same like iGoogle, Netvibes other organization provides such service with User customization, like SkyDeck [17] and Google Voice convergence. Skydeck is online Phone, which work to provide calls, text messages, voicemails and store contacts (on Skydeck.com [17]) User can search, read, and reply to your messages (by voice or by text) from Dashboard. Google Voice is a new service that ties all your phones together with one new number that rings them all. Using Dashboard SkyDesk provide own service as well combine Google Voice too.
8. Unlike converge, Dashwire [18], an Dashboard which work to Enabling Mobile Operators, as well as Device Makers and Retailers, to quickly and cost effectively deliver a new generation of customized consumer services on open mobile phone platforms. Using this Dashboard user receive Mobile Communication as well Web-interface together.

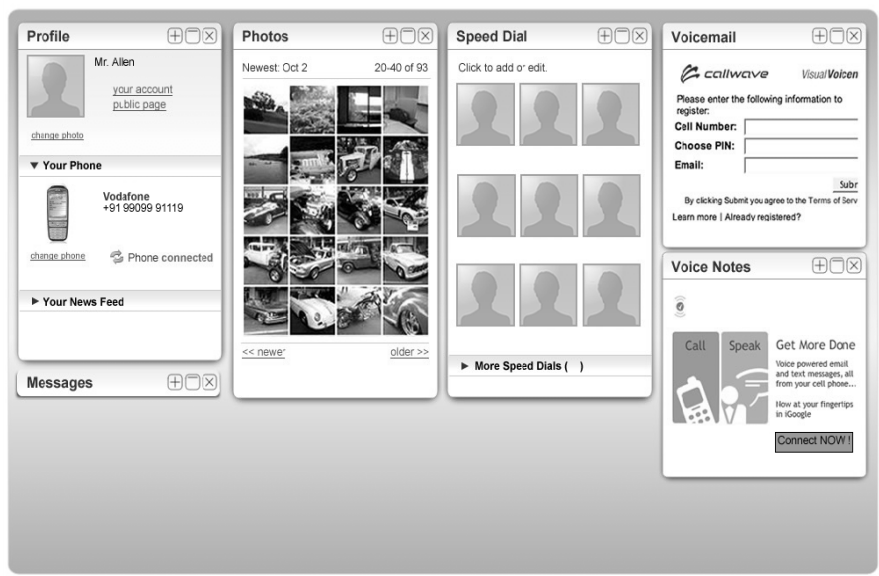

**Figure 3:** Sample Web Page unite Internet and telecom dashboard (Implemented for testing) [11]

## 5. Architectural descriptions

In this section we give a high-level architecture description of the proposed framework. Figure 3 displays a component overview of the framework. Components of our framework are categorized into two parts: the server side component and the client side component. Server side components of this architecture manage the persistent data such as user information with their preferences, service selection details, service configuration information and all such. For client side gadgets, all get organized and performs as per users' preference.

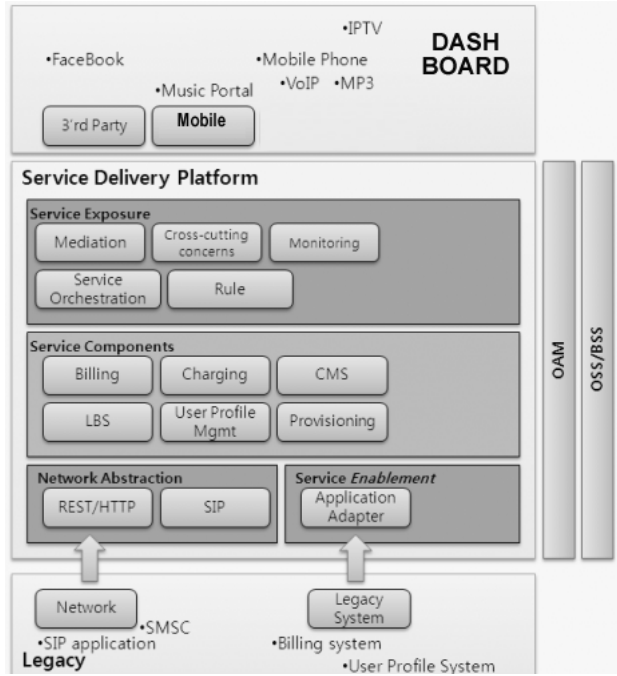

**Figure 4:** Architecture for Internet and telecom dashboard

This architecture works in manner to perform with authentication and after verifying the web server sends all gadgets data as per user preferences. Although all information must reveals with user preference manages and components / module. More briefly each module and their abs and all such flow user settings given at the. As per discussed in section 3 users' personalization is provided with Dashboard, which tightly bound to position and each setting with only user settings. Component data received from server and presented in specified module, all modules link up with AJAX technology with that easily updates module data no need to refresh whole web page. Any voice-video-data service using NGN also provided through this modules only.

## 6. Implementation issues

Our implementation of the dashboard is based on web technologies such as JavaScript, AJAX and PHP. With this dashboard we implemented a broad framework to pursue web feeds and their services aggregation. As per figure 4 the complete framework designed from End user client side to server includes service-managing tools like Billing/charging, session, user preference manager etc.

In Present architecture works from client side works in Web Browser with Java Script and AJAX, where the server side remain execute into a web server. Main issue with this Dashboard is web browser version. User works with Internet Explore and Firefox. But it is not possible all browser allows AJAX to run between multiple domains into a single web page. If browser not support instant data transfer from client to server it may cause late update of content into particular gadget.

To overcome this limit, we propose a proxy. The proxy works through same domain as the framework with the use of PHP, proxy enables and eases client side request towards server side. Some module may not able to constant render service request to server. But the proxy by pass this all requests and able to download to appropriate gadget. To this concern of security, all other AJAX security must protected outside their individual modules. This is known as module download controller component

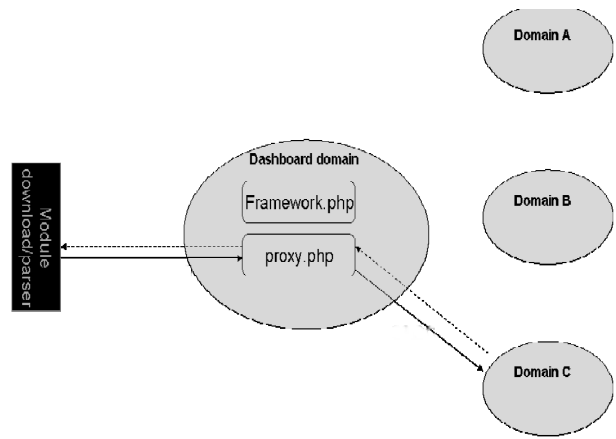

**Figure 5**: Security constraint of web browsers

In this proxy, server side works with PHP**.** The main parts are:

1. A proxy through which gadget gets loaded all requested services.
2. A component that manages users' database.
3. A component that manages the users' preferences database.

While Module download controller receives the response in form of XHTML, the requested service through the proxy and makes the necessary modifications on the module for modules independent.

> *e.g. http://webservice.com will be transformed to*
>
> *http://dashboardarea/proxy.php? url=http://webservice.com*

For providing browsing into a module, this also needs to change link of and URL to AJAX requests. Such modification also avoids reloading the whole page when a user clicks on any selected module links. Moreover, the follow necessary steps also considered for manage API,

1. E.g. if the developer of the module need to handle the unclose event,
2. Use the aggregator API; Add the following statement: ON_UNLOAD = handler

## 7. Conclusion

Aggregating different services in a single web page is a typical web2.0 and NGN way towards access and to use telecom services. Dashboard user easily configure his own choice, data over gadgets converged into dashboard. Once service made available over dashboard then after user will no longer user particular website, definitely will derived through dashboard only. "Innovative Web 2.0 companies are leveraging this viral environment to grow, unleashing the creativity of the developer community by allowing them to combine or "mashup" internal functionality or content with one or more other sources, using easy, standards-based drop and play tools, to create new value-added blended applications"

Adopting this approach, Mobile users able to access and use all his communication related services (like speed dial, calendar, visit Card) through a single web page. This is only an experimental observation but in Future work aims to implement over bulk user groups, for analyzing real time usage and Quality of Service. XHTML, XML is portable language but rather than SMIL 2.0 is also one alternate approach to consider for further implementation.

**References**
[1]     3GGP - "IP Multimedia Subsystem (IMS); Stage 2" - 3GPP TS 23.228 V6.6.0 (2004-06)
[2]     Nidal Nasser, Ming Shang, "Policy control framework for IP Multimedia Subsystem", Proceedings of the 2009 conference on Information Science, Technology and Applications, Pages: 53-58
[3]     Miikka Poikselka, Aki Niemi, Hisham Khartabil, Georg Mayer, "The IMS: IP Multimedia Concepts and Services," ISBN: 0470019069.
[4]     E. Thomas, Service-Oriented Architecture: A Field Guide to Integrating XML and Web Services. Prentice Hall, Upper Saddle River, NJ, 2004.
[5]     M. J. Yates and I. Boyd "The Parlay network API specification", BT Technology Journal, Volume 25, Numbers 3-4 / July, 2007, Pages 205-211

[6]     O'Reilly T. "What is Web 2.0. Design patterns and business models for the next generation of software." O'Reilly Media, 2005.
[7]     Linda Dailey Paulson, "Building Rich Web Applications with Ajax," Computer, vol. 38, no. 10, pp. 14-17, Oct., 2005
[8]     Zepeda, J.S. Chapa, S.V. "From Desktop Applications Towards Ajax Web Applications," Electrical and Electronics Engineering, pp 193-196, 5-7 Sept. 2007
[9]     Volker Hoyer, Marco Fischer, "Market Overview of Enterprise Mashup Tools", 6th International Conference Service-Oriented Computing – ICSOC 2008, Sydney, Australia, December 1-5, 2008. Page no. 708-721
[10]    O. Casquero, J. Portillo, R. Ovelar, J. Romo, and M.Benito, "iGoogle and gadgets as a platform for integrating institutional and external services", in: F. Wild, M. Kalz, and M. Palmér (Eds.), Mash-Up Personal Learning Environments. Proc. of 1st Workshop MUPPLE'08, Maastricht, 2008, pp. 37-41
[11]    C Kaar, "An introduction to Widgets with particular emphasis on Mobile Widgets," Computing, Oct. 2007.
[12]    Knights, M., "WEB 2.0 MySpace, Netvibes, Wikipedia, YouTube, Goowy... the web is moving from the personal to the social. In the process it is becoming increasingly complex, COMMUNICATIONS ENGINEER -IEEE 2007, VOL 5; ISSU 1, pages 30-35
[13]    Alison Lee, "Mobile Web Widgets:Enabler of Enterprise Mobility Work", 2nd Workshop on Mashups, Enterprise Mashups and Lightweight Composition on the Web (MEM 2009), Spain, Page No. 86-94
[14]    K Knightson, N Morita, T Towle, "NGN architecture: generic principles, functional architecture, and implementation", IEEE Communications Magazine, 2005
[15]    M Gomez, TP de Miguel, "Advanced IMS Multipoint Conference Management Using Web Services", - IEEE Communications Magazine, 2007
[16]    Niklas Blum, Thomas Magedanz, Horst Stein, "Service creation & delivery for SME based on SOA / IMS", Proceedings of the 2007 Workshop on Middleware for next-generation converged networks and applications table of contents, Page no. 311- 320
[17]    Nilanjan Banerjee, Koustuv Dasgupta, "Telecom mashups: enabling web 2.0 for telecom services", Proceedings of the 2nd international conference on Ubiquitous information management and communication, Suwon, Korea, Pages: 146-150
[18]    B.Falchuk, K.Sinkar, S.Loeb, A.Dutta, "Mobile Contextual Mashup Service for IMS", 2nd IEEE International Conference on Internet Multimedia Services Architecture and Applications (IMSAA-08), Bangalore, Dec. 2008
[19]    Dr. Kamaljit I. Lakhtaria, Dhinaharan Nagamalai, "Analyzing Web 2.0 Integration with Next Generation Networks for Services Rendering", The Third International Conference on Network Security & Applications (CNSA-2010), 2010, India, Page No. 581-591
[21]    Hiren Nagar, Billy B. L. Lim, "Mobile Computing With Web 2.0: Current State-of-The-Art, Issues & Challenges", *Journal of Computer & Information System VOL IX, No. 2, 2008, page no.523-529*
[22]    Boyera, Stéphane,. "Opportunities and Challenges of Web Technologies on Mobile Platform", http://www.w3.org/2007/08/sb_gitex/all.htm#(2), 2007